\documentclass[aps,pre,superscriptaddress]{revtex4}
\usepackage{latexsym} 
\usepackage{times} 
\usepackage{graphicx}

\usepackage[usenames,dvipsnames]{xcolor}

\usepackage{afterpage}

\usepackage{graphicx}
\usepackage{enumerate}
\usepackage{amsmath,amsthm,amscd,amssymb}
\usepackage[labelfont={bf}]{caption}

\newcommand{\change}[1]{ #1 }

\begin{document}

\title{ Spatial aggregation and the species-area relationship across scales }

\author{Jacopo~Grilli}
\affiliation{Department of Physics and Astronomy G. Galilei, Universit\`{a} di Padova, CNISM and INFN, via Marzolo 8, 35131 Padova, Italy}

\author{Sandro~Azaele}
\address{Institute of Integrative and Comparative Biology, University of Leeds, Miall Building, Leeds LS2 9JT, United Kingdom}

\author{Jayanth~R~Banavar}
\address{Department of Physics, University of Maryland, College Park, MD 20742, USA}

\author{Amos~Maritan}
\affiliation{Department of Physics and Astronomy G. Galilei, Universit\`{a} di Padova, CNISM and INFN, via Marzolo 8, 35131 Padova, Italy}


\begin{abstract}
There has been a considerable effort to understand and quantify the spatial distribution of species across different ecosystems. Relative species abundance (RSA), beta diversity and species area relationship (SAR) are among the most used macroecological measures to characterize plants communities in forests. In this article we introduce a simple phenomenological model based on Poisson cluster processes which allows us to exactly link RSA and beta diversity to SAR. The framework is spatially explicit and accounts for the spatial aggregation of conspecific individuals. Under the simplifying assumption of neutral theory, we derive an analytical expression for the SAR which reproduces tri-phasic behavior as sample area increases from local to continental scales, explaining how the tri-phasic behavior can be understood in terms of simple geometric arguments. We also find an expression for the endemic area relationship (EAR) and for the scaling of the RSA.
\end{abstract}

\maketitle

\section{Introduction}

\begin{figure}[tbp]
\centering
\includegraphics[width=0.6\textwidth]{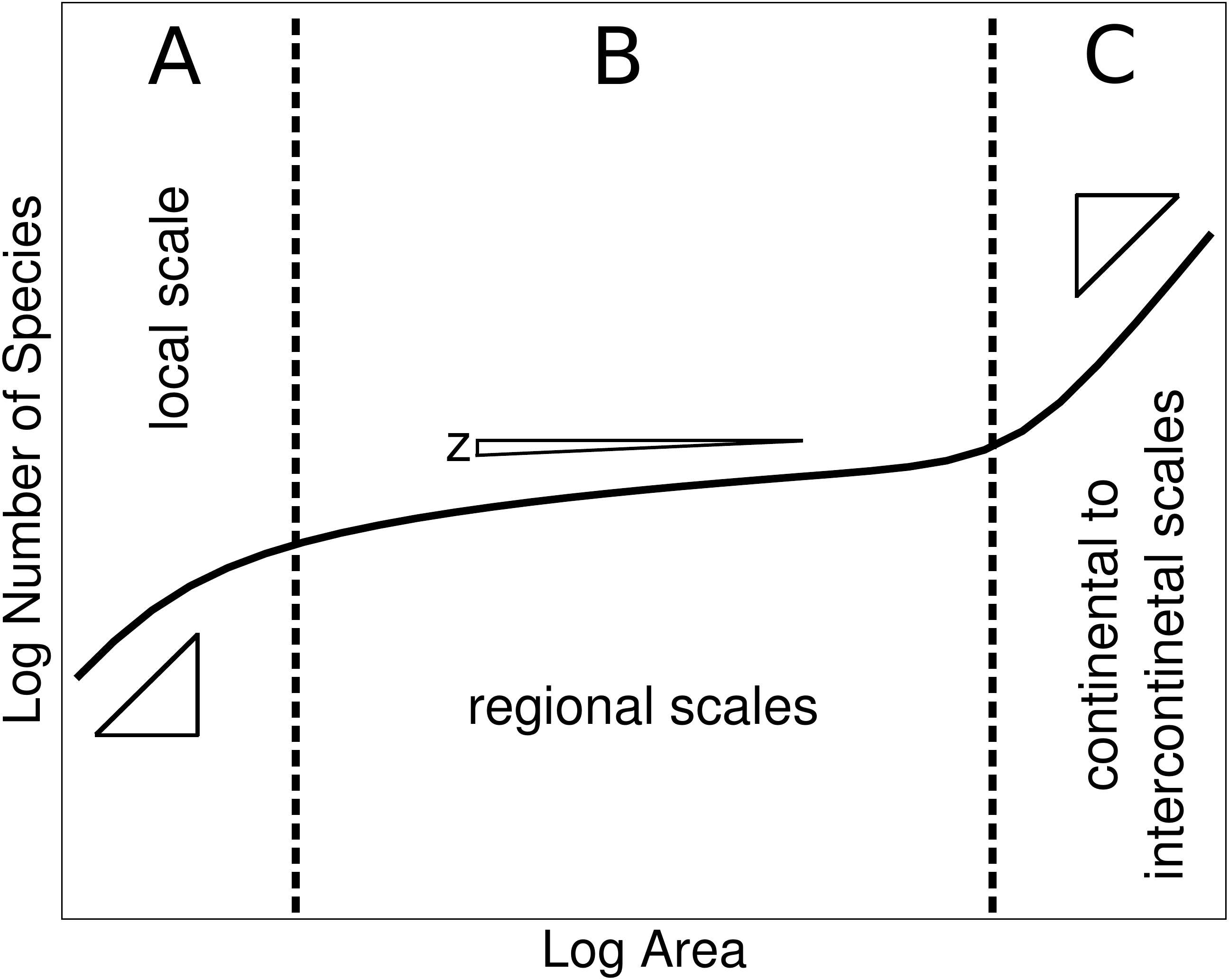}
  \caption{Qualitative shape of the Species-Area relationship~\citep{Hubbell2001a}.
	On local spatial scales (region A) the trend is steep.
	On intermediate spatial scales (region B) the slope decreases and the curve is well approximated by a power law with exponent $z$. Finally, on
	very large spatial scales (region C), the linear size of sampled areas is much greater than the correlation length of biogeographic
	processes, so that the majority of species are completely independent of each other.}
\label{fig:sar3shape}
\end{figure}

The relation between the mean number of different species observed within a given sampled area, i.e. the Species-Area relationship (SAR), is one of the most studied patterns in ecology and represents one of the simplest ways to characterize the biodiversity of a region. There is a considerable body of research~\citep{Arrh1921,Wilson67,May1975,Williamson88,HE1996,Storch07} showing that the curve of the SAR is a non-decreasing function whose slope depends on the sampled area and has a characteristic shape in a log-log plot (see Figure~\ref{fig:sar3shape}). This ``tri-phasic'' curve is relatively steeper at local and continental scales, but shallower at intermediate scales. This latter regime is typically described by a power-law $S \sim A^z$, even though there is no compelling theoretical reason to choose such a function. A wide range of models has been suggested in recent years to account for this shape: some of them are based on geometrical~\citep{Storch2008} or statistical considerations~\citep{GarciaMartin2006,Harte2009}, while others show that there are biological traits which can affect the shape of the SAR~\citep{Drakare2006}.

Simplified theoretical frameworks of population dynamics, such as the neutral theory of biodiversity~\citep{Hubbell2001a}, have made considerable progress in predicting several patterns at different spatial and temporal
scales~\citep{Chave2002,Volkov2003,Chave2004,Zillio2005,Alonso2006,Azaele2006,Volkov2007}, including the SAR~\citep{Rosindell2007}. Despite the simplicity of its core assumption, i.e. all individuals within a trophic level have the same probabilities to die or survive irrespective of the species which they belong to, the framework has provided a baseline expectation for a variety of patterns akin to those observed in empirical data. Thus, on the one hand it represents a powerful tool to investigate a series of underpinning mechanisms at the core of universal ecological behaviors; on the other, it questions more complex explanations for empirical patterns.  The original formulation~\citep{Hubbell2001a} and the majority of neutral models suggested later on have dealt with spatial features only implicitly \change{or the predictions are obtained via scaling relations~\cite{Zillio2008}}. Within such approaches the dispersal abilities of species are captured only approximately, although in an analytical tractable way. Spatially explicit models represent a substantial step towards a more realistic study of ecosystems, but present much greater theoretical challenges with respect to their implicit counterparts. In fact, spatial ecological measures such as the species area relationship crucially depend on the behavior of multiple points correlation functions, and any truncation would inevitably impair the predictions. As a consequence, one needs to solve the model in full generality, a task that is highly non trivial because stochastic theories defined on space often have stationary states for which detailed balance does \emph{not} hold. This condition ensures that at stationarity the probability to go from one configuration to another one is the same as the reversed transition~\cite{VanKampen1981,Zia2007}. Recently, O'Dwyer \& Green~\cite{O'Dwyer2010} have derived the SAR from a fully spatially explicit model by using field theoretical techniques. However, their findings were implicitly obtained under the assumption that the Detailed Balance is satisfied~\cite{DGans12}, a condition that is not correct for their model.

Within a neutral setting, we introduce a simple mechanism which is able to produce a tri-phasic
SAR and can be explained in simple geometrical terms. The model is based on the Poisson Cluster Processes~\citep{Cressie1993,Illian2008} and allows us to derive the SAR, the endemic area relationship and also the spatial scaling of the RSA.

\section{Emergent geometry of Poisson Cluster Processes}

Poisson Cluster Processes and Neyman-Scott processes~\citep{Cressie1993,Illian2008,Thomas1949,Neyman1958,Diggle1984} (PCP in the following) are a very general framework useful to analyze spatial ecological data and characterize population
aggregation~\citep{Plotkin,Morlon2008,Azaele2012}. These processes are quite simple and based on the assumption that individuals are spatially clumped in clusters. Specifically, the centers of clusters are distributed in space with a constant
density independent of each other. Each cluster is populated by a random number of individuals
(drawn from a given distribution) and the distance of each individual from the center of the
cluster is drawn from a given distribution (typically a Gaussian distribution with a certain
variance $\xi^2$).

We consider a simplified version of the PCP in a homogeneous landscape
of area $A_0$, assuming that:
\begin{enumerate}
\item Species are independent of each other. \label{item:indep}
\item The individuals of any species are distributed around a single center whose location is uniformly drawn within the landscape. \label{item:density}
\item The position of individuals with respect to the center is drawn from a given
distribution $\phi( \underline{r} )$, where $\underline{r}$ is the position with respect to the center. The distribution has a characteristic scale
$\xi$ above which it decreases exponentially.\label{item:corrlength}
\item The number of individuals per species are drawn from a given Relative Species Abundance (RSA) distribution $S_k(A_0)$. \label{item:rsa}
\end{enumerate}

The assumption in item 3 takes into account that individuals belonging to the same species are usually spatially aggregated (Plotkin 2002) -- we use a single cluster center (item 2) for the sake of simplicity. Here we do not focus on the biological mechanisms underlying conspecific spatial aggregation, but we account for it in a phenomenological fashion. Because we assume that species are independent (item 1) and every species is characterized by the same model parameters, the model is non-interacting and neutral as well.

The model is formulated as neutral, i.e. every species behaves in the same way. If neutrality holds and under the assumption of species independence,
we can consider simply one species at a time to calculate every quantity.
Within this model we can explicitly calculate  the SAR for a homogeneous and large landscape.
Under these hypotheses we obtain the species-area relationship simply from
the probability of finding at least one individual in a given sub-region of area $A$
by~\citep{O'Dwyer2010}:
\begin{equation}
\displaystyle
S(A|A_0) = S_{tot}(A_0) \sum_{k=1}^\infty P_k(A|A_0) = S_{tot}(A_0) \big[ 1 - P_0(A|A_0) \big] \ ,
\label{eq:SAR-def}
\end{equation}
where $S_{tot}(A_0)$ is the total number of available species in the whole system with area $A_0$ (i.e. $\sum_{k=0}^\infty S_k(A_0)$),
while $P_k(A|A_0)$ is the probability of finding exactly $k$ individuals of a given species in the sub-region of
area $A$ and has the following expression (see~\ref{sec:supp-P_k(A|A_0)})
\begin{equation}
\displaystyle
P_k(A|A_0) = \int_0^\infty d \lambda p(\lambda)
\frac{1}{A_0} \int_{A_0} d^2 \underline{r} \frac{\big[\lambda \int_{A(\underline{r})}
d^2 \underline{r}' \phi(\underline{r}') \big]^k}{k!}
e^{-\lambda \int_{A(\underline{r})}
d^2 \underline{r}' \phi(\underline{r}')} \ ,
\label{eq:P_0(A_0)}
\end{equation}
where $A(\underline{r})$ is a region of area $A$ centered at the point $\underline{r}$.
The distribution $p(\lambda)$ is strictly related to the RSA, $S_k(A_0)$, implicitly defined by the following equation
\begin{equation}
\displaystyle
S_k(A_0) = S_{tot}(A_0) \int_0^\infty d\lambda p(\lambda) \frac{\lambda^k e^{-\lambda}}{k!} \ .
\label{eq:RSAtot}
\end{equation}
Interestingly, the equation~\ref{eq:P_0(A_0)} reduces to the random placement model~\citep{RandPlace} in its mean field version,
i.e. by considering $\phi(\underline{r})$ to be constant. 
On the other hand, it is possible to relate the quantity $\phi(\underline{r})$
to the two point correlation function (see~\ref{suppsec:rhodep}).
The correlation function is proportional to the $\beta$-diversity (which is a well known
measurable quantity in real systems~\citep{Condit2002}). Specifically, we obtain
the following relation
\begin{equation}
\displaystyle
G_2(\underline{r})= \big< \lambda^2 \big> \int_{A_0} d^2 \underline{y} \phi(\underline{y}) \phi(\underline{y}-\underline{r}) \ ,
\label{eq:corfun}
\end{equation}
where $\big< \lambda^2 \big> = \int_0^\infty d \lambda  \lambda^2  p(\lambda)$.
Thus we can directly obtain an expression for $\phi(\underline{r})$ when the correlation function.
By applying the Fourier transform, it is possible to invert equation~\ref{eq:corfun}, obtaining $\widehat{\phi}(\underline{p}) \propto \sqrt{ \widehat{G}_2(\underline{p})}$ (where $\widehat{\phi}(\underline{p})$
is the Fourier transform of $\phi(\underline{r})$, see~\ref{suppsec:rhodep}).
The formula in Eq. \ref{eq:corfun}, with an appropriate choice of $\phi(\underline{r})$, has the same structure obtained with different models~\citep{Condit2002,Zillio2005}.

\section{Results}

\subsection{Species Area Relationship}

\begin{figure}[tbp]
\centering
  \includegraphics[width=0.9\textwidth]{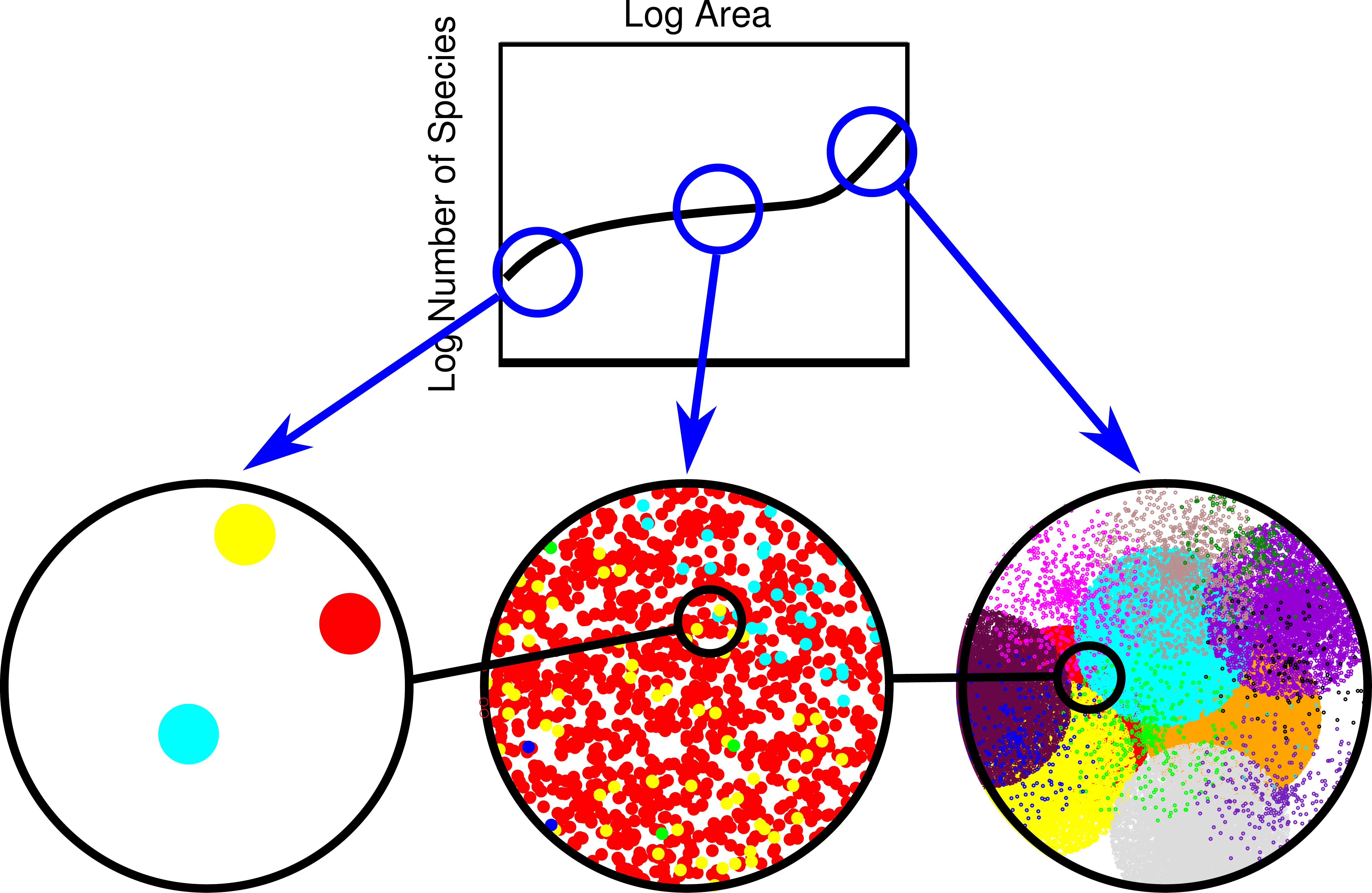}
  \caption{This figure shows the mechanism which produces the tri-phasic SAR as
	explained by the model. Different colors indicate different species. At large spatial scales the sampled areas are larger than the typical one 		 occupied by a given species: this produces the linear
	scaling observed at large areas. When we observe the system at intermediate scales,
	the distribution of individuals follows a non-trivial spatial organization
	which corresponds to power-law-like behavior. Instead, at very small
	scales, on average every individual belongs to a different species, thus making the scaling with the area linear. This shows that the tri-phasic SAR can be understood in terms of very general geometric considerations. \change{The figure at large scales is obtained, for graphical reasons, in a regime of relative small $s$, which introduces strong fluctuations in the density of individuals. The density of individuals is constant for reasonable values of $s$.}}
\label{fig:geometry}
\end{figure}

The final expression of the Species-Area Relationship is obtained by substituting equation~\ref{eq:P_0(A_0)}
into equation~\ref{eq:SAR-def} and taking the limit $A_0 \to \infty$ in the spirit of~\citep{Rosindell2007}.
We find (see~\ref{sec:supp-limitA_0}) that the average number of
species found within a sampled area is equal to
\begin{equation}
\displaystyle
S(A) = s_{tot} \int d^2 \underline{r} \Bigl[ 1 -
\int_0^\infty d \lambda p(\lambda) e^{- \lambda \int_{A(\underline{r})} d^2 \underline{r}' \phi(\underline{r}') }
\Bigr] \ .
\label{eq:SAR-final}
\end{equation}
The quantity $s_{tot}$ is obtained from the limit $\lim_{A_0 \to \infty} S_{tot}(A_0)/A_0$ and it has the interpretation
of an effective density of species. In~\ref{sec:supp-limitA_0}, we show that this quantity is well defined, i.e. the limit
does exist and is different from zero. Note that in this way we have obtained an analytic expression
for the SAR if we are given the RSA and the pair correlation function. The model generates the spatial aggregation of individuals in a very simple way and without resorting to any explicit biological mechanism (see Figure~\ref{fig:geometry}), and therefore the emergent spatial distribution could potentially describe spatial features of species with very different traits, dispersal abilities or habitat preferences. Thus, the SAR in equation~\ref{eq:SAR-final} is more the result of basic geometrical features than the effect of underlying biological mechanisms. This consideration is important especially when one tries to infer the effects of fundamental mechanisms simply by comparing empirical data to analytical curves obtained from more complex models.

We can extract some general information from equation~\ref{eq:SAR-final} independent of the specific form of the RSA and the correlation function.
. Because $\xi$ is a correlation length and characterizes the spatial scale over which a species is distributed, from dimensional analysis (see~\ref{suppeq:DimensionalAnalis}) we have that $S(A) = s_{tot} A f(A/\xi^2)$.
We can study the SAR for small and large areas (which can be obtained as an expansion
for $A \gg \xi^2$ and $A \ll \xi^2$). The small area expansion gives, regardless of the
choice of the RSA or the spatial distribution, the following result (see~\ref{sec:supp-expanzion})
\begin{equation}
\displaystyle
S(A) \sim \big< \rho \big> A = N(A) \ ,
\label{eq:SAR-scale_small}
\end{equation}
where $\big< \rho \big>$ is the density of individuals and $N(A)$ is the number of
individuals in the area $A$. This is an expected result: when we sample small areas, the majority of
sampled individuals belong to different species (and thus the number of species grows linearly with the number of individuals). This result ($S \sim N$) is valid for all areas when $\phi(r) = \text{cost}$ and corresponds to the Random Placement model~\citep{RandPlace} in the limit
of large $A_0$.
For large areas (i.e. areas much larger than the one given by the typical correlation scale), we obtain
\begin{equation}
\displaystyle
S(A) \sim \ s_{tot} \Bigl( 1 - \int_0^{\infty} d \lambda p(\lambda) e^{-\lambda} \Bigr) = s A  \ ,
\label{eq:SAR-scale_large}
\end{equation}
where $s$ is defined as the average density of observable species (i.e. with at least one individual in the whole landscape).
At large spatial scales the mean number of species grows linearly with the sampled area and the spatial aggregation of individuals is no longer important, only the total density of species $s$ matters.

\begin{figure}[tbp]
\centering
  \includegraphics[width=0.9\textwidth]{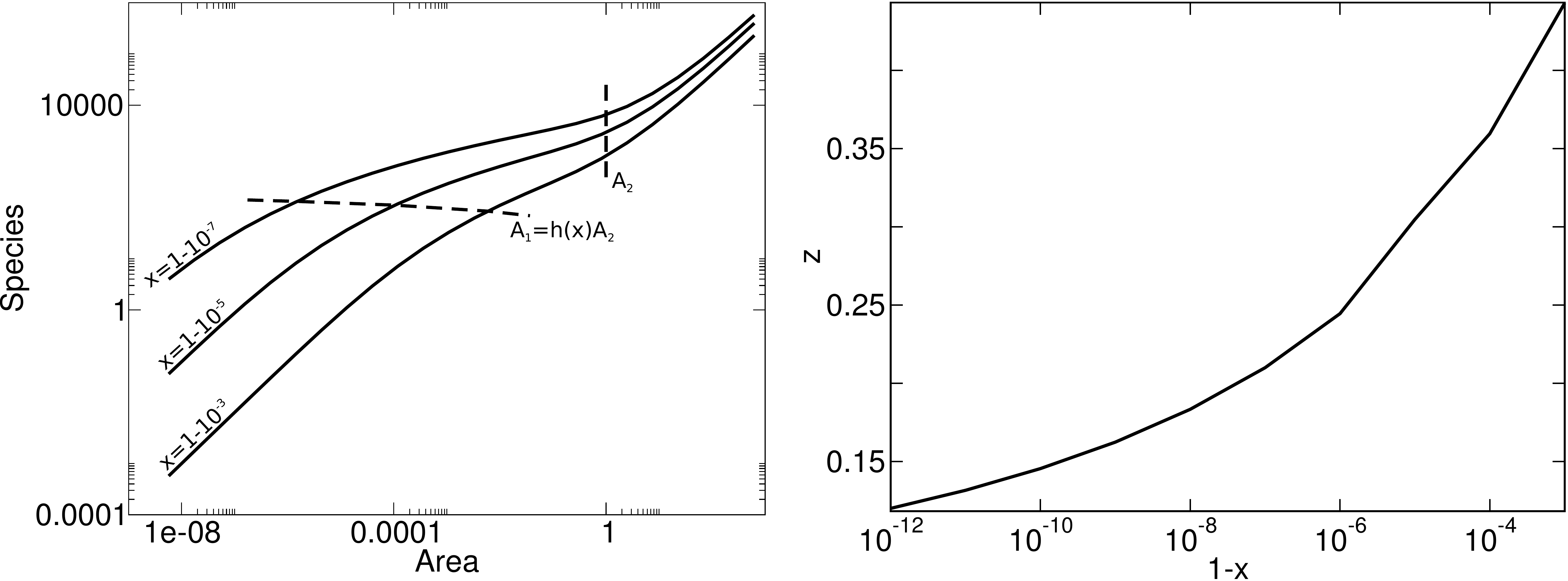}
  \caption{The Species-Area relationship and the exponent $z$. The left panel shows the SAR obtained for different values of $x $ (solid lines) where we have set $\theta=1$. This choice is justified because the qualitative behavior of $S(A)$ and the exponent $z$ are independent of $\theta$. The area is measured in units of $A_2$, the area at which the linear behavior of the SAR sets in. The dashed black segments represent the two scales $A_1$ and $A_2$ obtained in equation~\ref{eq:first_scale} which separate the different regimes. The right panel shows the values of the exponent $z$ (obtained at the inflection point) for different values of $x$. It spans the observed values for reasonable values of $x$.}
\label{fig:sar}
\end{figure}

Now we focus on specific forms of the RSA and the correlation function in one idealized example.
We consider the Fisher log-series for the RSA~\citep{Fisher1943a,Hubbell2001a}
(i.e. $S_k = \theta x^k / k$) and the Bessel function $K_0$ for the correlation function~\citep{Condit2002}.
The corresponding choice for $p(\lambda)$ is an appropriate limit of the Gamma distribution (curiously
the Fisher log-series was firstly introduced by Fisher~\citep{Fisher1943a} exactly via $p(\lambda)$),
while for $\phi(\underline{r})$ we obtain (see~\ref{sec:supp-K_0Fisher})
\begin{equation}
\displaystyle
\phi(\underline{r}) = \frac{\exp(- ||\underline{r}||/\xi)}{\xi ||\underline{r}||}
\label{eq:rho_r}
\end{equation}
By substituting this expression in equation~\ref{eq:SAR-final} we obtain
\begin{equation}
\displaystyle
S(A) = \theta \int d\underline{r} \log \Bigl( \frac{1 - x \big(1 - I(A,\underline{r}) \big)}{1-x} \Bigr)  \ .
\label{eq:SAR-fisher}
\end{equation}
where $I(A,\underline{r})=\int_{A(\underline{r})} d \underline{r}' \phi(\underline{r}')$. In general, the integral in Eq.\ref{eq:SAR-fisher} does not have a closed form, however it can be easily evaluated numerically and the result is shown in Figure~\ref{fig:sar} for
$\phi(\underline{r})$ given in equation~\ref{eq:rho_r}.

The SAR shows a linear growth at small as well as large scales
as predicted by the general consideration above
and an approximate power-law at intermediate regions.
The scale between the power-law trend and the large area linear growth is totally determined
by the shape and characteristic scale of the correlation function (i.e. of the $\beta$-diversity). For example, if we consider $\phi(r)$ equal to zero
outside a circular region with radius $\xi$ and constant inside, the scale will be equal to $A_2=\pi \xi^2$.
In our case, where we have used $K_0$ as the correlation function, we obtain $A_2= \xi^2 16 \pi $ (see~\ref{suppsec:scales}).
In Figure~\ref{fig:sar} we plot the result in units of this area, showing that the scale does not depend
on $x$. The scale $A_1$ between the rapid growth at small scales and the power-law behavior depends on the RSA through the parameter $x$.
We observe a rapid growth at small areas because we are sampling individuals of different species, this trend
starts to bend when we collect more individuals of the same species. This happens at a scale equal to the typical
distance between conspecific individuals, i.e. the scale $A_1$ is the average area occupied by one individual of a given species.
In~\ref{suppsec:scales} we calculate this scale to be
\begin{equation}
\displaystyle
A_1 = h(x) A_2 \equiv (1-x) \frac{-x-\log(1-x)}{x^2} A_2 \ .
\label{eq:first_scale}
\end{equation}
In Figure~\ref{fig:sar} this quantity is plotted for the SAR with different values of $x$.
We find that the SAR shows a linear trend with a slope equal to density of
individuals $\big< \rho \big>$ for areas $A< A_1=h(x)A_2$, a power-law trend $S\sim A^z$ for $A_1 < A < A_2$ and
a linear growth at scales $A> A_2$ where the proportionality constant is equal to the average density of species $s$.

At intermediate scales the derivative $d\log S(A)/d\log A$ varies slowly, so the behavior of the SAR can be well approximated by a power-law if the exponent $z$ is defined as the slope at the inflection point, i.e. the minimum of $d\log S(A)/d\log A$. We show the result in the right panel of Figure~\ref{fig:sar}. The exponent $z$, in this version of the model, depends only on the parameter $x$ and ranges between $0.15$ and $0.4$, which is the range of observed values see e.g.~\citep{Hubbell2001a}.
The parameter $x$ is the parameter of the Fisher log-series, which is assumed to be the RSA of the entire system.
By using the relations which relate the speciation rate and the density of individuals~\citep{Volkov2003},
we obtain that, for reasonable values of the speciation rate $\nu$, $1-x \sim s \nu/\big<\rho\big>$.
The model predicts a value of the exponent $x$ between $0.15$ and $0.4$ for reasonable values of $1-x$
between $10^{-3}$ and $10^{-9}$~\citep{Condit2002}.

The model allows one to calculate not only the SAR, but also the probability to find $k$ species within an area $A$.
Under the hypotheses of neutrality and of the absence of interactions, the probability $1-P_0(A|A_0)$ to find a given species 
in a certain sub-region of area $A$ is independent on the other species. Due to the absence of interactions,
the joint probabilities to find a given set of species factorize and then the probability to find $k$ species in a sub-region of area $A$
will be a Binomial distribution
\begin{equation}
\displaystyle
P_k^S(A|A_0) = \binom{S_{tot}(A_0)}{k} \big( 1-P_0(A|A_0) \big)^k \big( P_0(A|A_0) \big)^{S(A_0)-k} \ .
\label{peq:Sprob_fin}
\end{equation}
In the limit of $A_0 \to \infty$, $S_{tot}(A_0)$ tends to infinite while $1-P_0(A|A_0)$ tends to $1$ with a
finite product  and thus the distribution in the large $A_0$ limit turns to be a Poisson
distribution with average $S(A)$
\begin{equation}
\displaystyle
P_k^S(A) = \frac{ \big( S(A) \big)^k }{k!} \exp(-S(A)) \ .
\label{eq:Sprob}
\end{equation}
Therefore in the large total area limit the probability to find $k$ species in a sub-region of area $A$ is a Poisson distribution
with average $S(A)$.
This is a prediction that could be simply tested with empirical data. The result is not specific of our model
but is generally valid under the non-interacting assumption. Thus this represents an interesting and practical
way to measure  the macroscopic effect of the interactions between species at different scales.

\subsection{Endemic Area Relationship}

While the SAR is defined as the average number of species in an area $A$, the Endemic Area Relationship (EAR)~\cite{He2011}
is the average number of species whose individuals are completely contained in an area $A$. This quantity has a
fundamental importance in ecology, because gives an estimation of the number of immediate
extinction due to a loss of space (further extinctions  might take later).
Within our framework we can obtain an expression EAR and its relation with the SAR.

\begin{figure}[tbp]
\centering
  \includegraphics[width=0.9\textwidth]{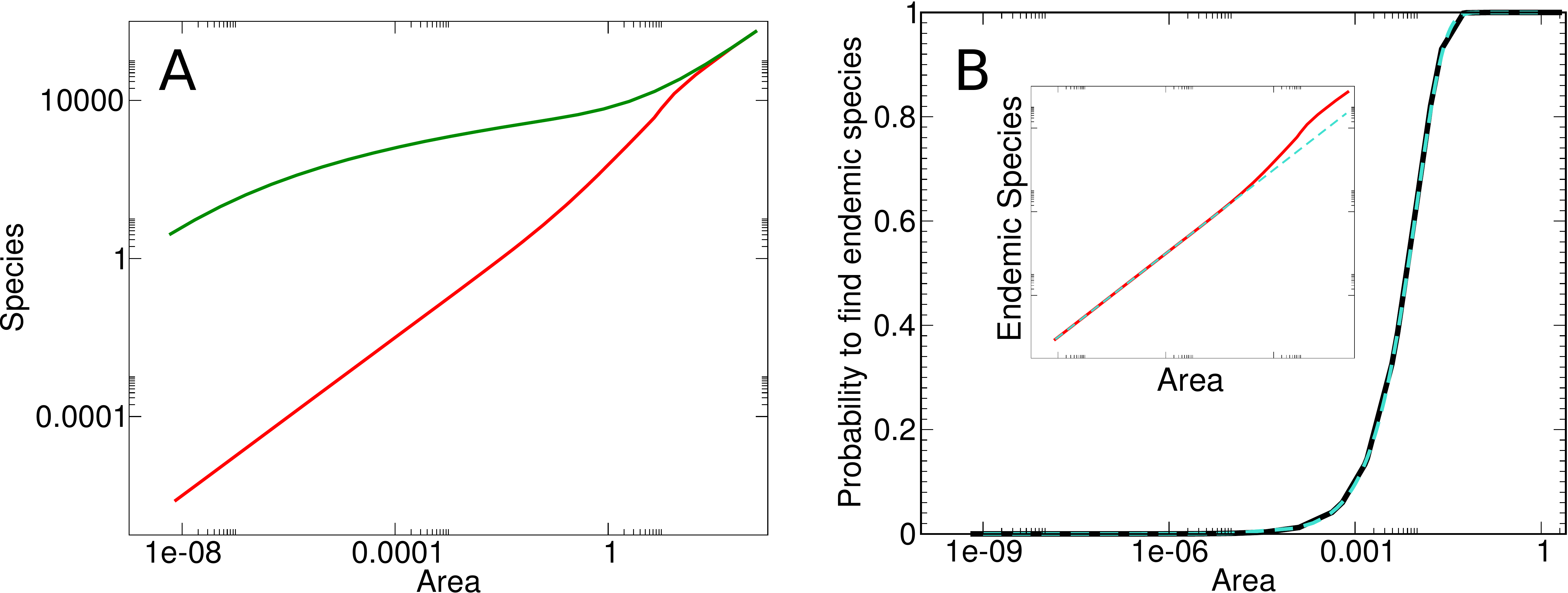}
  \caption{Endemic-Area Relationship. In the Figure A we compare the SAR (green curve) with the EAR (red curve),
	obtained respectively from equation~\ref{eq:SAR-fisher} and~\ref{eq:Ear-fisher}. The curve are plotted
	for $x=1-10^{-7}$ and $\theta=1$. In Figure B we show the probability to find at least
	one endemic species (see equation~\ref{eq:Ear-prob}). The black curve is obtained by integrating equation~\ref{eq:Ear-fisher},
	while the turquoise dotted curve is obtained with the approximation of the equation~\ref{eq:Ear-approx}.
	The figure inside is a comparison between the EAR and the approximation at small scales of equation~\ref{eq:Ear-approx}.
	The area unit are the same as in figure~\ref{fig:sar}.}
\label{fig:ear}
\end{figure}

The general formula for the EAR
can be obtained by calculating the number of species with zero abundance outside a sub-region
of area $A$ (see~\ref{suppsec:EAR})
\begin{equation}
\displaystyle
E(A) = s_{tot} \int d^2 \underline{z} \int_0^\infty d \lambda p(\lambda) e^{- \lambda} \big[
e^{ \lambda \int_{A(\underline{z})} d^2 \underline{r} \phi(\underline{r})  } -1 \big]
 \ .
\label{eq:EAR-final}
\end{equation}
This expression depends on the distributions $p(\lambda)$
and $\phi(r)$ which are related to the RSA of the entire system and to the $\beta$-diversity.
The EAR corresponding to the case analyzed for the SAR is
\begin{equation}
\displaystyle
E(A) = - \theta \int d\underline{z} \log \Bigl( 1 - x I(A,\underline{z}) \Bigr)  \ .
\label{eq:Ear-fisher}
\end{equation}
This expression is compared to the scaling of the SAR (see equation~\ref{eq:SAR-fisher})
in Figure~\ref{fig:ear}.
Interestingly, the EAR seems to be linear up to the correlation length. The EAR becomes quite similar to
the SAR for length scales larger than the correlation length, because we are considering areas much larger than the typical
space occupied by a species. In Figure~\ref{fig:ear} we observe that the EAR shows a linear trend at small scales, by expanding
equation~\ref{eq:Ear-fisher} we obtain
\begin{equation}
\displaystyle
E(A) \sim  \theta x A   \ ,
\label{eq:Ear-approx}
\end{equation}
if $A \ll A_2$. This approximation is equivalent to the result of the random placement~\cite{RandPlace}.
Note that the trend of the EAR depends weakly on the value of $x$ when it assumes the empirical values
which are typically close to $1$. On the contrary its depends on the biodiversity parameter $\theta$. This approximation,
as shown in Figure~\ref{fig:ear}, is valid, for values of $x$ close to $1$,
also at the scales at which the SAR shows the power-law trend (which are the most interesting scales from theoretical
point of view based on the experience gained in statistical mechanics of continuum transition characterized by power-law behavior and
universality).

We can also calculate, as done for the SAR, the probability distribution of EAR,
defined as the probability that in an area $A$ there are $k$
endemic species. By using the same arguments used for SAR we demonstrate in the~\ref{suppsec:EAR} that the EAR follows
a Poisson distribution with average $E(A)$.
It is interesting to study the probability to find at least one endemic
species $P^{EAR}_e(A)$, which is distributed as
\begin{equation}
\displaystyle
P^{EAR}_e(A) = 1 - P^{EAR}_0(A) = 1 - e^{-E(A)}  \ ,
\label{eq:Ear-prob}
\end{equation}
where $P^{EAR}_0(A)$ is the probability that any endemic species is found in an area $A$. The
plot of this quantity is shown in Figure~\ref{fig:ear}. We observe that this probability has a non trivial
scaling with a rapid increase at the scale at which $E(A)$ approaches to one. Therefore it exists a typical
scale over which we observe endemic species. Our framework allows to determine this scale. As shown in Figure~\ref{fig:ear}
the scaling of the probability is well approximated, at the interesting scales, by substituting the expansion of the EAR
at small scales. By using the expansion of equation~\ref{eq:Ear-approx} we can calculate the typical area $A_c$ at which
the probability of equation~\ref{eq:Ear-prob} becomes equal to $1/2$
\begin{equation}
\displaystyle
A_c = \frac{\log(2)}{\theta x}  \ .
\label{eq:Ear-scalehalf}
\end{equation}
This expression is valid only if $A_c \ll A_2$, because it follows from the expansion of equation~\ref{eq:Ear-approx}.
But this is the interesting case, because, for the typical ecological application (e.g. to have an estimation of the
extinction debt), it is important to know the EAR behavior at small scales. Note in Figure~\ref{fig:ear} that,
due to the shape of the EAR, the linear approximation for the EAR is a lower bound of its real value (i.e.
the expression of equation~\ref{eq:Ear-approx}, which is a good approximation at small scales, is always lower than the real value
at bigger scales). This fact simply implies that the real value of the area $A_c$ (the area at which the probability to find an endemic
species is equal to $1/2$) is always lower than the value of equation~\ref{eq:Ear-scalehalf}, which is a good approximation at small scales.

\subsection{Relative Species Abundance}
\label{sec:rsa}

In this section we obtain an expression for the RSA restricted to a given sub-region. This quantity is defined as the number of species $S_k$
with a certain individual abundance $k$. To obtain an expression for the SAR and the EAR, we have postulated a form of
the RSA in the whole landscape $S_k(A_0)$. Starting from this input, within our framework, we can obtain an expression for
the RSA $S_k(A)$ in a sub-region of area $A$.

The general formula for the RSA restricted to a sub-region is obtained in~\ref{sec:supp-P_k(A|A_0)} and, in the limit
of large $A_0$, turns to be
\begin{equation}
\displaystyle
S_k(A) = s_{tot} \int d^2 \underline{z}
\int_0^\infty d \lambda p(\lambda)
\frac{ \Bigl[ \lambda \int_{A(\underline{z})} d^2 \underline{r} \phi(\underline{r}) \Bigr]^k}{k!}
\exp \Bigl( - \lambda \int_{A(\underline{z})} d^2 \underline{r} \phi(\underline{r}) \Bigr) \ .
\label{eq:RSA-subreg}
\end{equation}
Note that this expression is consistent with our expression for the SAR:
by summing over $k$ we obtain equation~\ref{eq:SAR-final}. As done for the SAR and the EAR, we study the case in which
the RSA of the entire system is a Fisher log series, obtaining
\begin{equation}
\displaystyle
S_k(A) = \frac{ \theta }{k} \int d^2 \underline{z}
\Bigl( \frac{x I(A,\underline{z})}{1-x\big( 1- I(A,\underline{z}) \big)} \Bigr)^k \ .
\label{eq:RSA-fisher}
\end{equation}

\begin{figure}[tbp]
\centering
  \includegraphics[width=0.9\textwidth]{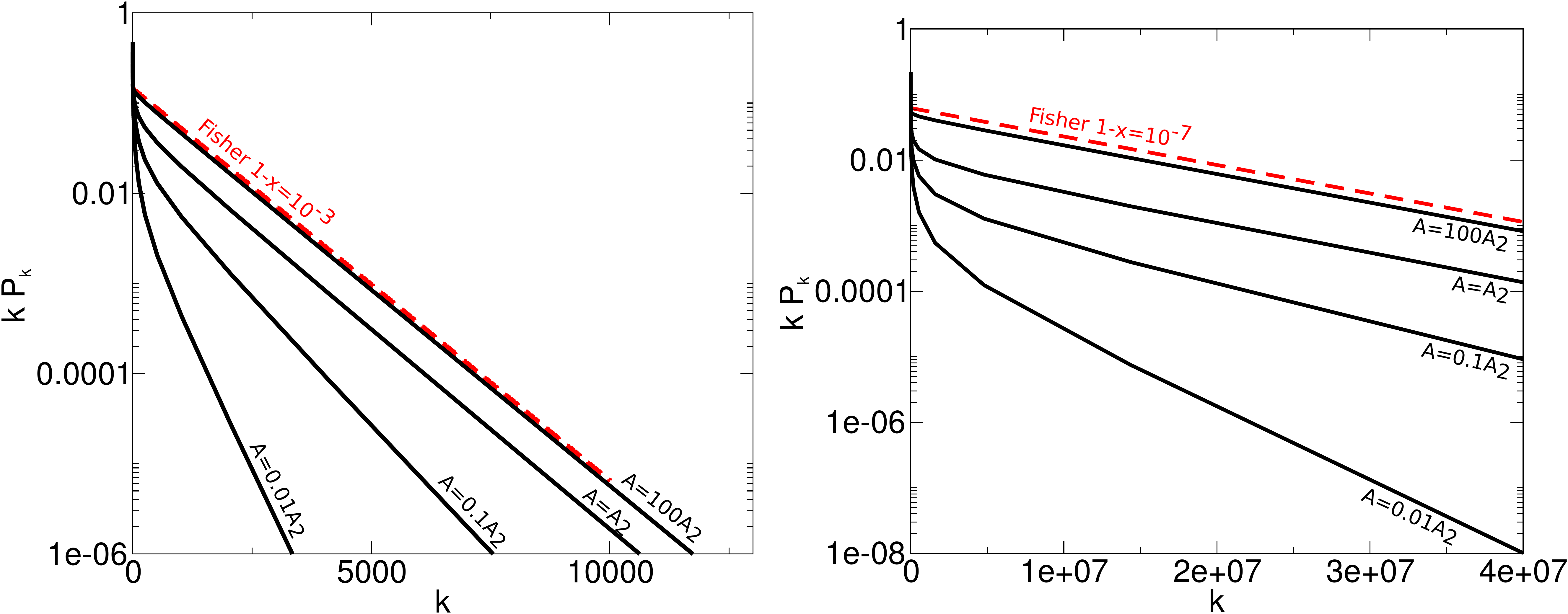}
  \caption{Species Abundance Distribution at different scales. In this figure we plot $k P^{NRSA}_k(A)$
	at different scales. The quantity $P^{NRSA}_k(A)$ is defined for $k >0$ and is obtained as the ratio between
	the RSA of equation~\ref{eq:RSA-fisher} and the SAR of equation~\ref{eq:SAR-fisher}.
	The left figure is for a value of $x$ equal to $1-10^{-3}$, while the right one
	is obtained for $x=1-10^{-7}$. The dotted red line is
	the NRSA at largest scale (which is in our model a Fisher log-series), while the black continuous
	lines are the NRSA at different scales. As the area decrease the slope of the curve
	(which is directly related to x) decreases.
	Notably this effective x
	decreases with the observed area, as observed in empirical systems, since the smaller the system the smaller is the average population per species 	and so a rapid decreases of the RSA at large population, i.e. at smaller value of $x$}
\label{fig:rsa}
\end{figure}

In order to compare different length scales we do not use directly the RSA, which depends extensively on the observed area,
instead we study the behavior of the Normalized Relative Species Abundance (NRSA). This quantity is defined as the probability
$P^{NRSA}_k$ that an observed species has a certain number of individuals and it could be obtained by normalizing the RSA to one (i.e. by
dividing it with the SAR). In our case the NRSA of the entire system is a Fisher log-series, i.e. $P^{NRSA}_k(A_0)=x^k/(-k\log(1-x))$.
In Figure~\ref{fig:rsa} is shown the behavior of this probability at different length scale. We observe different behaviors
at different scales. Very interestingly if we measure the parameter $x$ via the large $k$ behavior of $P^{NRSA}_k(A)$, we
find an effective parameter $x_{eff}(A)$ which depends on the area observed $A$ and is equal to $x$ in the limit $A \gg A_2$.
Notably this effective $x$ decreases with the observed area, as observed in empirical systems.

\section{Discussion}

In this article we have introduced a simplified version of the Poisson Cluster Processes apt to describe large homogeneous landscapes. This is not a microscopically based process, but describes the aggregation of individuals starting from simple phenomenological considerations. Within this framework, we have shown how one can relate the SAR to the beta-diversity~\citep{Condit2002}
and to the RSA under simple and general assumptions. Secondly, we have obtained the tri-phasic SAR
and identified how the exponent of the approximated power-law depends on the parameters of the RSA (e.g. the demographical parameter or the speciation
rate). Finally we have obtained a formula for the EAR and an expression for the RSA at different scales.

In order to disentangle different sources of information within species-area patterns, we first need to understand how general the assumptions are that can generate the observed patterns. If the qualitative shape of a curve can be captured by simple mathematical considerations, then it seems likely that ecological aspects drive finer and more quantitative details of the curve, although alternative explanations may hold as well. Our work shows that the tri-phasic shape of the SAR is a very general pattern that emerges under simple and general geometrical considerations. Specifically, sampling individuals on local scales and the spatial aggregation of conspecific individuals on larger yet finite scales (note that this defines the characteristic length scale for $\beta$-diversity) produce the two bending points in the SAR, eventually making the curve tri-phasic. Accordingly, the pattern is rather qualitatively insensitive to the implementation of specific ecological mechanisms, and thus it is not surprising that models based on very different hypotheses~\citep{Rosindell2007,DeAguiar2009} can account for tri-phasic SARs.
Within this context it is possible to relate the prediction of the SAR with the form of the $\beta$-diversity. The effects of inter- and intra-specific interaction, spatial heterogeneity and species' traits are important when dealing with the fine details of the curve and should be taken into account when a precise prediction is necessary. These mechanisms could influence, in a non trivial way, the final SAR curve and the value of the exponent $z$.

We have shown that the exponent $z$ (measured as the inflection point of the SAR curve) depends on the demographic parameter of the
RSA distribution. Although the measured
values of $z$ reflect a more complicated dynamics which produces complex spatial patterns
at intermediate scales, we find, that for realistic values of the demographical parameter $x$, the
exponent $z$ spans the empirically observed values.
We have also obtained a way to infer the typical scales
at which the power-law trend is observable. It is well known that the measure of the exponent depends
on the scale we observe it~\citep{Drakare2006}. However, we have shown that the range of scales where the measure of the exponent $z$ is relatively more reliable is directly linked to the correlation length.

In this work our results are expressed in terms of the demographical parameter $x$, 
the parameter characterizing the RSA at largest scale, i.e. the Fisher log-series.
As shown in section~\ref{sec:rsa} $x$ is not the demographic parameter of the system at every scale. When we observe the RSA at a smaller spatial scale, we obtain a different distribution.
However an effective demographical parameter can be defined at each spatial scale in terms of the decay of the RSA tail at the same spatial scale.
We get that this effective demographical parameter is an increasing function of the area as is also empirically observed.

The assumption of non-interacting species within the same trophic level makes it possible to calculate from the SAR the probability to find a given number of different species in a certain area. We found that this probability is a Poisson distribution (in the limit of a large landscape $A_0$, while it turns out to be a Binomial distribution when a finite $A_0$ is considered). This quantity could be
measured in available datasets and it represents a powerful way
to identify the spatial scales over which neutrality is a good approximation and at what scales the interaction
becomes macroscopically observable.

Within our framework it is also possible to calculate an analytical expression for the endemic-area relationship (the number of species which are completely contained in a certain area) and its distribution. Interestingly, the EAR
scales linearly at small scales. We obtained the linear scaling as expansion for small areas which is equivalent
to the random placement model~\citep{RandPlace}.
In a recent work~\citep{He2011} is shown that the random placement describes with a very good approximation
the behavior of the EAR in different data set. These data set refers to systems with a finite total area $A_0$,
therefore our model (which is valid for $A \ll A_0$) is not a good candidate to describe those systems.
On the other hand, our framework is able to give an explanation of why the random placement works in a good way to describe
the EAR, but is not able to reproduce the trend of the SAR. In fact we have shown that the the random placement is a good
approximation for the EAR at scales lower than $A_2$ (the typical space occupied by a species),
whereas it describes the trend of SAR only below the scale $A_1$ (the typical area occupied by a single individual per species).
Our model provide also an expression for the distribution of the EAR, allowing to calculate the typical area at which there
is a non negligible probability to find an endemic species.

\change{The model could also be used to test and to compare the validity
of the predictions obtained via scaling relations.
For instance, it is possible to show that the scaling relations,
which describe the behavior at local scales well~\cite{Zillio2008},
are also valid for our model in the limit of small areas (where the random placement is recovered).
This is not true for larger areas, where it could be interesting to study the
appearance of new simple relations between the observed quantities.
}

The model we propose can be extended in several different ways. Firstly, it would be useful to
study how the SAR curve varies according to different sampling methods.
It is known that the measured SAR depends on the sampling scheme (e.g., nested
vs. independent)~\citep{Drakare2006} and on the geometry of the sampled area~\citep{Kunin1997}.
It is not trivial to understand whether these differences are, in principle, simply quantified
by geometrical considerations or whether they hide some biologically relevant aspects.
Finally, it would be interesting to introduce non-neutral characteristics and inter-species interaction.

\change{The model we have introduced does not follow from any intrinsic dynamics but it captures,
in a simplified and meaningful way, many different processes acting on different spatial and temporal scales.
We have shown that regardless of any specific dynamics, the patterns observed in empirical
studies, especially at large spatial scales, can be explained on the basis of quite general and simple processes.
It would be interesting to incorporate simple dynamics into the model to assess how the spatial patterns are affected.}

We have proposed an analytically solvable model based on minimal assumptions. It allows us to calculate
explicitly the SAR on an infinite landscape, and also the EAR and the scaling of the RSA. Although this approach neglects important characteristics
of ecosystems, it allows us to understand the necessary (geometrical or biological) mechanisms at the core of
the observed macroecological patterns and therefore to quantify the relative importance of the neglected effects.

\newpage


\renewcommand{\thesection}{Appendix \Alph{section}}
\setcounter{figure}{0}
\setcounter{equation}{0}
\setcounter{table}{0}
\setcounter{section}{0}

\renewcommand{\figurename}{Supplementary Figure}
\renewcommand{\tablename}{Supplementary Table}
\renewcommand{\thefigure}{A\arabic{figure}}
\renewcommand{\thetable}{A\arabic{table}}
\renewcommand{\theequation}{A\arabic{equation}}


\section{Calculation of $P_k(A|A_0)$}
\label{sec:supp-P_k(A|A_0)}

In this section we want to calculate the probability to find exactly $k$ individuals of a given species
in a sample area $A$. This quantity is directly related to the SAR.
We sketch this calculation starting from the hypotheses written
in the main text. As explained before, the model we propose is a simplified version of the
Poisson Cluster Processes~\citep{Cressie1993,Illian2008} to which we refer for a more extensive and rigorous
discussion.

The model is neutral and non-interacting. This assumption makes possible to obtain an analytical expression for
the SAR, because it implies that we can consider
one species at a time.
 
A simple and intuitive way to perform this calculation is to consider discrete space,
write the probability we are interest for, and calculate the final result in the continuum limit.
In order to distinguish the quantities defined in the continuum and on a lattice, we indicate
a quantity with a $\tilde{\cdot}$ when it is considered in discrete space.

Consider a homogeneous and isotropic lattice $\Lambda$ with periodic boundary conditions.
A site of this lattice is identified by a vector $\underline{r}$.
We assume that a single site could be empty or occupied by a single individual.
We know from item~\ref{item:density} of our assumptions that the individuals of a species are
distributed in a single cluster centered in a
point of space $\underline{x}$. We define $\tilde{p_1}(\underline{r}|\underline{x})$ as the probability that we find an
individual in a point $\underline{r}$ given $\underline{x}$ to be the position of the center of the cluster,
whereas the probability
that we find the site $\underline{r}$ empty will be simply
$\tilde{p_0}(\underline{r}|\underline{x})=1-\tilde{p_1}(\underline{r}|\underline{x})$.

Consider a set of sites $A_{\underline{z}} = \{ \underline{r}_1, \dots, \underline{r}_{|A_{\underline{z}}|} \}$ which
has a cardinality $|A_{\underline{z}}|$. We identify this set by labeling it with a point of the lattice $\underline{z}$.
We can calculate, by using the quantities we have just introduced, the probability to find $k$ sites occupied and the others empty.
It becomes
\begin{equation}
\displaystyle
\tilde{P}_k(A_{\underline{z}}|\underline{x}) =  \frac{|A_{\underline{z}}|!}{k!(|A_{\underline{z}}|-k)!} \sum_{ (\underline{r}_1,\dots,\underline{r}_k) \in A_{\underline{z}}}
\Bigl[ \prod_{ \underline{r} \in \{\underline{r}_1,\dots,\underline{r}_k\} } p_1(\underline{r}|\underline{x}) \Bigr]
\Bigl[ \prod_{ \underline{r} \in A_{\underline{z}} \setminus \{\underline{r}_1,\dots,\underline{r}_k\} }
\big( 1 - p_1(\underline{r}|\underline{x} ) \big) \Bigr] \ .
\label{suppeq:ksitesProb}
\end{equation}
This expression defines the probability $\tilde{P}_k(A_{\underline{z}}|\underline{x})$
to find $k$ individuals when we are observing a set of sites $A_{\underline{z}}$, when the cluster of individuals is centered in a
point $\underline{x}$. This expression is valid without imposing any constraint on the set $A_{\underline{z}}$, but we want
to interpret it as an area centered in a point of the space $\underline{z}$, when the continuum limit will be performed.
Thus we consider $A_{\underline{z}}$ as a set of $|A_{\underline{z}}|$ sites distributed around the point $\underline{z}$ in such a way that this
set converges in the continuum limit to a region $A(\underline{z})$ with an area $A$ centered in $\underline{z}$.
We are in principle not interested in the dependence on the location of the sample and on the location of the
cluster center.
Thus we have to average $\tilde{P}_k(A_{\underline{z}}|\underline{x})$ over possible choices of $\underline{x}$
and $\underline{z}$. We obtain the following expression
\begin{equation}
\displaystyle
\tilde{P}_k(A) = \frac{1}{|\Lambda|} \sum_{\underline{z} \in \Lambda}
\frac{|A_{\underline{z}}|!}{k!(|A_{\underline{z}}|-k)!} \sum_{ (\underline{r}_1,\dots,\underline{r}_k) \in A_{\underline{z}}}
\Bigl[ \prod_{ \underline{r} \in \{\underline{r}_1,\dots,\underline{r}_k\} } p_1(\underline{r}) \Bigr]
\Bigl[ \prod_{ \underline{r} \in A_{\underline{z}} \setminus \{\underline{r}_1,\dots,\underline{r}_k\} }
\big( 1 - p_1(\underline{r}) \big) \Bigr] \ ,
\label{suppeq:ksitesProb3}
\end{equation}
where $p_1(\underline{r})=p_1(\underline{r}|\underline{0})$.

Considering the definition of $p_1(\underline{r})$, the average number of individual placed around
a cluster center will be 
$\lambda=\sum_{\underline{r} \in \Lambda} p_1(\underline{r})$.
We introduce a new quantity $\tilde{\phi}(\underline{r})$ defined by the following
relation $p_1(\underline{r}) = \lambda \tilde{\phi}(\underline{r})$.
Note that $\tilde{\phi}(\underline{r})$ carries all the spatial information about $p_1(\underline{r})$.

To obtain the expressions in the continuum limit, we have to introduce a finite site spacing, define the scaling of the
quantities respect to it and calculate the limit of vanishing site spacing.
By performing this calculation in two dimensions we obtain
\begin{equation}
\displaystyle
P^\lambda_k(A|A_0) = \frac{1}{A_0} \int_{A_0} d^2 \underline{z}
\frac{ \Bigl[ \lambda \int_{A(\underline{z})} d^2 \underline{r} \phi(\underline{r}) \Bigr]^k}{k!}
\exp \Bigl( - \lambda \int_{A(\underline{z})} d^2 \underline{r} \phi(\underline{r}) \Bigr) \ ,
\label{suppeq:ksitesProbContLambdaA}
\end{equation}
where $A_0$ is the area of the whole landscape, $A(\underline{z})$ is a region (e.g. a circle) centered
in the point $\underline{z}$ and $\phi(\underline{r})$ is the continuum limit of $\tilde{\phi}(\underline{r})$.

We would like to introduce in equation~\ref{suppeq:ksitesProbContLambdaA} our knowledge of
the RSA $S_k(A_0)$ (see item~\ref{item:rsa} of our assumptions).
The knowledge of the RSA gives us an information about the probability to find $k$ individuals in the
whole landscape (usually called Species Abundance Distribution, SAD): starting from the RSA we know that
\begin{equation}
\displaystyle
P_k(A_0) = \frac{S_k(A_0)}{S_{tot}(A_0)} \ ,
\label{suppeq:ksitesProbContLambdaA}
\end{equation}
where $S_{tot}(A_0)$ is the total number of available species, which is given by $\sum_{k=0}^\infty S_k(A_0)$
and $P_k(A_0)$ is the SAD.
We want that the probability calculated with our model match the one obtained starting from the SAD when the
whole landscape is considered. The expression calculated with our model in equation~\ref{suppeq:ksitesProbContLambdaA}
depends on a parameter $\lambda$. We assume this parameter to be a random variable distributed in the
interval $(0,\infty)$ accordingly to a probability distribution function $p(\lambda)$. This distribution
$p(\lambda)$ will be auto-consistently determined by imposing the matching between the model and the SAD
when the whole system is considered. This procedure does not hide any particular ecological meaning, it is only
a trick to perform the calculation and to impose the condition on the RSA.

The probability obtained in equation~\ref{suppeq:ksitesProbContLambdaA} evaluated in an area $A=A_0$
becomes a Poisson distribution with average $\lambda$
\begin{equation}
\displaystyle
P^\lambda_k(A_0) = \frac{  \lambda^k}{k!} e^{-\lambda}  \ .
\label{suppeq:ksitesProbContLambdaTot}
\end{equation}
By introducing a distribution $p(\lambda)$ we obtain in the most general case
\begin{equation}
\displaystyle
P_k(A_0) = \int_0^\infty d \lambda p(\lambda) \frac{  \lambda^k}{k!} e^{-\lambda} := \frac{S_k(A_0)}{S_{tot}(A_0)} \ .
\label{suppeq:ksitesRSA}
\end{equation}
This expression defines $p(\lambda)$ in terms of the $RSA$.
This equation is valid for $k \ge 0$, i.e. we are $S_{tot}$ counts even the species with zero abundance in the whole landscape
(it is not a directly measurable quantity). In other words the probability $P_0(A_0)$ is generally different from zero.
The total number of observable species (i.e. the species
with at least one individual in the whole landscape) will be given by
\begin{equation}
\displaystyle
S(A_0) = S_{tot}(A_0) (1- \int_0^\infty d \lambda p(\lambda) e^{-\lambda}) \ .
\label{suppeq:Stot}
\end{equation}

By introducing $p(\lambda)$ in equation~\ref{suppeq:ksitesProbContLambdaA}, we finally obtain
\begin{equation}
\displaystyle
 S_k(A|A_0) := S_{tot}(A_0) P_k(A|A_0) 
 = \frac{S_{tot}(A_0)}{A_0} \int_{A_0} d^2 \underline{z}
\int_0^\infty d \lambda p(\lambda)
\frac{ \Bigl[ \lambda \int_{A(\underline{z})} d^2 \underline{r} \phi(\underline{r}) \Bigr]^k}{k!}
\exp \Bigl( - \lambda \int_{A(\underline{z})} d^2 \underline{r} \phi(\underline{r}) \Bigr) \ , 
\label{suppeq:ksitesProbContLambda}
\end{equation}
and the number of species turns to be
\begin{equation}
\displaystyle
S(A|A_0) := \sum_{k=1}^\infty S_k(A|A_0) = \frac{S_{tot}(A_0)}{A_0} \int_{A_0} d^2 \underline{z}
\int_0^\infty d \lambda p(\lambda)
\Bigl[ 1 -
\exp \Bigl( - \lambda \int_{A(\underline{z})} d^2 \underline{r} \phi(\underline{r}) \Bigr) \Bigr] \ .
\label{suppeq:SARareacond}
\end{equation}

\section{Dependence on $ p(\lambda) $ and $ \phi(r) $}
\label{suppsec:rhodep}

Equation~\ref{suppeq:ksitesProbContLambda} depends only on two functions: $p(\lambda)$ and $\phi(\underline{r})$.
These two functions are respectively related to the distribution of individuals in species and to the distribution of
individuals in space.

The probability distribution function $p(\lambda)$ is directly related to the Relative Species Abundance.
The function $\phi(\underline{r})$ was instead introduced as related to the probability that
a site was or not occupied by one individual. Starting from the definition of the model,
we observe that the two point correlation function
is equal to
\begin{equation}
\displaystyle
G(\underline{r})= \big< \lambda^2 \big> \int_{A_0} d^2 \underline{y} \phi(\underline{y}) \phi(\underline{y}-\underline{r}) \ ,
\label{suppeq:2CorFunCont}
\end{equation}
where $\big< \lambda^2 \big> = \int_0^\infty d \lambda  \lambda^2  p(\lambda)$.
By applying the Fourier transform, it is possible to invert this expression obtaining
\begin{equation}
\displaystyle
\widehat{\rho}(p) = \sqrt {\frac{\widehat{G}(p)}{ \big< \lambda^2 \big> } } \ .
\label{suppeq:2CorFunCont-fourier}
\end{equation}
This expression gives us a direct way to infer from data a form of the $\phi(\underline{r})$ starting
from the correlation function (which has the same functional dependence of the $\beta$-diversity).
Note that, due to the normalization condition of $\phi(r)$, it is sufficient to know the functional dependence of the
correlation function (or of the $\beta$-diversity) to obtain the exact expression of $\phi(r)$. An example of this calculation is
shown in section~\ref{sec:supp-K_0Fisher}.

\section{Limit $ A_0 \to \infty $}
\label{sec:supp-limitA_0}

We are interested in calculating the following limit
\begin{equation}
\displaystyle
S(A) :=  \lim_{A_0 \to \infty}  S(A|A_0) \ .
\label{suppeq:SAR-limit}
\end{equation}
In order to perform this limit, we have to know how  $S_{tot}(A_0)$ scales with $A_0$ when we consider the limit
of large $A_0$. We affirm that the total number
of species scale as
\begin{equation}
\displaystyle
S_{tot}(A_0) \sim s_{tot} A_0 \ \ \text{if $A_0 \to \infty$} \ .
\label{suppeq:Stot-scal}
\end{equation}
This scaling is not an assumption, instead it is a consequence of the fact that the total number of individuals scale
with the area in the large area limit whereas the number of individual of a single species remains constant for sufficient
large areas.
Note that, due to equation~\ref{suppeq:Stot}, even $S(A_0)$ follows a linear scaling:
\begin{equation}
\displaystyle
S(A_0) \sim s A_0 \ \ \text{if $A_0 \to \infty$} \ ,
\label{suppeq:Stot-scal2}
\end{equation}
where $s$ is related to $s_{tot}$ via equation~\ref{suppeq:Stot}, i.e.
\begin{equation}
\displaystyle
s = s_{tot} \Bigl( 1- \int_0^\infty d \lambda p(\lambda) e^{-\lambda} \Bigr) = s_{tot} \big( 1 - P_0(A_0) \big)  \ .
\label{suppeq:s_vs_stot}
\end{equation}

By substituting the scaling of $S(A_0)$ and the sum over $k$ in equation~\ref{suppeq:SARareacond},
we obtain the following expression
\begin{equation}
\displaystyle
S(A) = s_{tot} \int d^2 \underline{z} \Bigl[ 1 -
\int_0^\infty d \lambda p(\lambda) e^{- \lambda \int_{A(\underline{z})} d^2 \underline{r} \phi(\underline{r}) }
\Bigr] \ ,
\label{suppeq:SAR-limit}
\end{equation}
which is our central result.

\section{Dimesional analysis}
\label{suppeq:DimensionalAnalis}

The equation~\ref{suppeq:SAR-limit} depends at least on three parameters: the density of species $s_{tot}$,
the parameter of the RSA (there is at least a single parameter which appears in the distribution $p(\lambda)$)
and the correlation length $\xi$ (which appears in $\phi(r)$).
The parameter $s_{tot}$ (which is not directly measurable, because it
represent the density of available species) could by related to the density of observable species $s$
by equation~\ref{suppeq:s_vs_stot}.
It is possible to determine the functional form of the SAR, by using the dimensional analysis.
The SAR is a function of $A$, which has the dimension of an area. The parameter $s$ is a density and
thus it has the dimension of an inverse of area, while $\xi$ is a length.
The parameter $s$ appears as a multiple of the entire expression leading to the dimensionalless result:
\begin{equation}
\displaystyle
S(A) = s A f \Bigl( \frac{A}{\xi^2} \Bigr) \ .
\label{suppeq:SAR-final2}
\end{equation}
The function $f$ depends also on the dimensionless parameter appearing in the RSA.

Note that if we did not consider the limit $A_0 \to \infty$ (i.e. we were interested in finite-size
scaling) the SAR would also depend on the size of the system $A_0$ and thus the functional dependence would be
more complicate.

\section{Expansion of SAR for small and large areas}
\label{sec:supp-expanzion}

The starting point to perform the expansions at small and large areas is the equation~\ref{eq:SAR-final} of the main text:
\begin{equation}
\displaystyle
S(A) = s_{tot} \int d^2 \underline{z} \Bigl[ 1 -
\int_0^\infty d \lambda p(\lambda) e^{- \lambda \int_{A(\underline{z})} d^D \underline{r} \phi(\underline{r}) }
\Bigr] \ .
\label{suppeq:SAR-final}
\end{equation}
As written above, this equation depends at least on three parameters and has the form
written in equation~\ref{suppeq:SAR-final2}.
In order to calculate the limit of small or large areas we have to evaluate
the previous expression for small or large ratios $ A/\xi^2$.

\textbf{Small area expansion.} When we consider small areas the integral
$\int_{A(\underline{z})} d^D \underline{r} \phi(\underline{r})$ tends to
$ A \phi(\underline{z})$. Expanding the exponential of equation~\ref{suppeq:SAR-final} we obtain
\begin{equation}
\displaystyle
S(A) \sim s_{tot} A \int_0^\infty d \lambda \lambda p(\lambda) = s_{tot} A \big< \lambda \big> = \big< \rho \big> A   \text{ \ \ if $ A \ll \xi^2$} \ .
\label{suppeq:SAR-smallA}
\end{equation}
Note that $\big< \lambda \big>$ is equal to the average number of individuals per species $\big< k \big>$ (see equation~\ref{suppeq:ksitesRSA}),
and thus $s_{tot} \big< \lambda \big>$ is equal to the average density of individuals $\big< \rho \big>$.

\textbf{Large area expansion.} Consider the integral
$\int_{A(\underline{z})} d^D \underline{r} \phi(\underline{r})$
for large areas. We know that $\phi(\underline{r})$ is a function which
decreases sufficiently rapidly for large areas, with a typical
scale $\xi$. Thus for large areas the integral could be well approximated by
the characteristic function $\chi_{A(\underline{0})}(\underline{z})$ (which is equal to
$1$ if $\underline{z}$ belongs to the region $A(\underline{0})$ and it is zero otherwise). We obtain
\begin{equation}
\displaystyle
S(A)  \sim  s_{tot} \int d^2 \underline{z} \Bigl[ 1 - \int_0^\infty d \lambda p(\lambda)
e^{-\lambda \chi_{A(\underline{0})} (\underline{z}) } \Bigr] = s_{tot} \big( 1 - P_0(A_0) \big)  = s A
  \text{ \ \ if $ A \gg \xi^2$} \ ,
\label{suppeq:SAR-largeA}
\end{equation}
where $s$ is the density of the species we observe in the entire system and is related to $s_{tot}$
via equation~\ref{suppeq:s_vs_stot}.

\section{A choice for $\phi(r)$ and $p(\lambda)$}
\label{sec:supp-K_0Fisher}

One of the most known form for the Relative Species Abundance is the Fisher log-series~\citep{Fisher1943a},
which is defined as
\begin{equation}
\displaystyle
S_k(A_0) = \tilde{\theta} \frac{x^k}{k} \ \text{if $k \ge 1$} \ ,
\label{suppeq:RSA-fisher}
\end{equation}
where $\tilde{\theta}>0$ and $x \in (0,1)$ are the two parameters of the distribution.
The total number of observable species will be
\begin{equation}
\displaystyle
S(A_0) = \sum_{k=1}^\infty S_k(A_0) =  - \tilde{\theta} \log(1-x)  \ .
\label{suppeq:SARtotal-fisher}
\end{equation}
We have shown is section~\ref{sec:supp-limitA_0} that the number
of observable species in the entire system scales linearly with $A_0$ if $A_0$ in sufficiently large.
We assume that for large $A_0$, $\theta \sim A_0$ whereas $x$ does not depend on it. This assumption
respects the requested scaling properties of $S(A_0)$ and it is in agreement with the microscopic interpretation
of the Fisher log-series (e.g. via birth-death process).
We define
\begin{equation}
\displaystyle
\theta := \lim_{A_0 \to \infty} \frac{ \tilde{\theta} }{A_0} \ .
\label{suppeq:RSA-fisher}
\end{equation}

The Fisher log-series was obtained, in the original derivation~\citep{Fisher1943a}, as
an appropriate limit of a convolution between a Gamma distribution and
a Poisson distribution
\begin{equation}
\displaystyle
S_k(A_0) = \lim_{\epsilon \to 0} S(\epsilon) \int_0^{\infty} d \lambda
\frac{e^{-\lambda/\delta} \lambda^{\epsilon-1}}{\Gamma(\epsilon) \delta^{\epsilon} } \frac{\lambda^k e^{-\lambda}}{k!}
:=\tilde{\theta} \frac{x^k}{k} \ .
\label{suppeq:RSA-fisherLim}
\end{equation}
The parameter $\tilde{\theta}$ is defined as the limit of $S(\epsilon)/\Gamma(\epsilon)$ for $\epsilon \to 0$,
whereas $x$ is defined as $\delta/(1+\delta)$.
We can see that equation~\ref{suppeq:RSA-fisherLim}
give us a recipe to choose the function $p(\lambda)$, because the RSA is exactly written in the same form of
equation~\ref{suppeq:ksitesRSA}. Thus in order to impose the Fisher log-series as the RSA for the entire landscape,
we have to choose $p(\lambda)$ as a appopriate limit of the Gamma distribution.

We can obtain an explicit expression for the SAR by substituting our choice of $p(\lambda)$. For a finite area
$A$ we obtain
\begin{equation}
\displaystyle
S_k(A|A_0) = \frac{1}{A_0} \int_{A_0} d^2 \underline{z} \lim_{\epsilon \to 0} S(\epsilon) \int_0^{\infty} d \lambda
\frac{e^{-\lambda/\delta} \lambda^{\epsilon-1}}{\Gamma(\epsilon) \delta^{\epsilon} } \frac{ (\lambda I(A,\underline{z}))^k }{k!}e^{-{\lambda I(A,\underline{z})}} \ .
\label{suppeq:RSA-fisherLim2}
\end{equation}
where $I(\underline{z},A)$ is defined as
\begin{equation}
\displaystyle
I(A,\underline{z})= \int_{A(\underline{z})} \phi({\underline{r}}) d^2\underline{r} \ .
\label{suppeq:Irho1}
\end{equation}

Performing the integral in equation~\ref{suppeq:RSA-fisherLim2}, taking the limit $\epsilon \to 0$ and the limit
$A_0 \to \infty$ and summing over $k$ from $1$ to $\infty$ we finally obtain the following expression for the SAR
\begin{equation}
\displaystyle
S(A) = \theta \int d\underline{z} \log \Bigl(  \frac{1 - x ( 1 - I(A, \underline{z} ) ) }{1-x} \Bigr) =
s \int d\underline{z} \frac{ \log \big(  \frac{1 - x ( 1 - I(A, \underline{z} ) ) }{1-x} \big)}{-\log(1-x)}  \ .
\label{suppeq:SAR-fisherA}
\end{equation}
Note that this expression when expanded for large and small area, follows the scaling obtained in
section~\ref{sec:supp-expanzion} as expected.

To obtain a tractable expression we have also to specify a recipe for the function $\phi(\underline{r})$. As demonstrated
above this function
could be related to the two point correlation function (or the $\beta$-diversity) by equation~\ref{suppeq:2CorFunCont}.
The two point empirical correlation function could be for example fitted by a Bessel function $K_0(r/\xi)$~\citep{Condit2002}
Following the procedure sketched in section~\ref{suppsec:rhodep}, we can obtain a functional form for $\phi(\underline{r})$
by calculating the Fourier transform of the two point correlation function, which for the choice of the Bessel function
turns to be
\begin{equation}
\displaystyle
\widehat{G}(\underline{p}) \propto \frac{1}{1+\xi^2 \underline{p}^2} \ ,
\label{suppeq:K_0Fourier}
\end{equation}
by taking its square root and by applying the Fourier anti-transform, we finally obtain 
\begin{equation}
\displaystyle
\phi(\underline{r}) = \frac{e^{-||\underline{r}||/\xi}}{\xi ||\underline{r}|| 2 \pi} \ .
\label{suppeq:rhoK0}
\end{equation}
In this expression the proportionality constant
was fixed by imposing the normalization condition $\int \phi(\underline{r}) d \underline{r} =1$ (see section~\ref{suppsec:rhodep}).

Thus with the choice for $\phi(\underline{r})$ expressed above, the integral $I(\underline{z},A)$
becomes
\begin{equation}
\displaystyle
I(A,\underline{z})=
\int \Theta(||\underline{r}-\underline{z}|| - R)
\frac{e^{-||\underline{r}||/\xi}}{ ||\underline{r}|| \xi} d^2\underline{r} \ ,
\label{suppeq:Irho}
\end{equation}
where we are considering a circular region $A(\underline{z})$ with an area $A=\pi R^2$.

\section{Scales}
\label{suppsec:scales}

The tri-phasic SAR, as shown in figure~\ref{fig:sar3shape}, seems to have two separate length scales $A_1$
and $A_2$. The first one separates the the linear trend at low scales with the power-law region, the second
one is the boundary between the power-low intermediate region and the linear trend at large scales. We show in this section
that our model give an expression for both the scales starting from only one length scale $\xi$ (the correlation
length).

Observing the figure~\ref{fig:geometry}, we can understand the mechanism which produces the observed pattern. The scale
$A_2$ above which we obtain the linear scaling is the typical area occupied by a species: above it we have sampled the
entire population of a single species. This scale depends only on $\xi$ and on the form of the
correlation function (see Figure~\ref{fig:sar}).

The first scale $A_1$ is determined by the typical minimum distance between two conspecific individuals (i.e. the average
distance between one individual and the nearest conspecific): below this
length scale, the sampled individuals belong to different species and thus the scaling is linear, above it the curve
starts to bend down because we are sampling multiple individuals of the same species.
This quantity could be well estimated from the RSA as the average of the reciprocal of the density (calculated in the
area where the species live), which gives the typical area occupied by only one individual of a given species.
Note that the distance between one individual and its nearest conspecific is well defined only if the species we are considering
has at least two individuals. 
Let us pick an individual at random (chosen between the individuals belonging to the species with a population of at least two individuals).
It will belong to a species with $k$ individuals. The portion of area in which this individual is the only one belonging to its species
will be well approximated by $A_2/k$. Let us pick an individual, the probability that it belongs to a species with $k$ individuals is
proportional to $k P_k$. We thus have to average this quantity with the probability to pick an individual of a species with
a total number of individuals equal to $k$ restricted to the condition to have at least two individuals (i.e. $k P_k/(\sum_{k \ge 2}k P_k)$).
We obtain
\begin{equation}
\displaystyle
A_1 = \sum_{k=2}^\infty \frac{A_2}{k} \frac{k S_k(A_0)}{\sum_{k=2}^\infty k S_k(A_0)}  =\frac{\sum_{k \ge 2} S_k(A_0)}{\sum_{k \ge 2} k S_k(A_0)} A_2 \ .
\label{suppeq:A-1}
\end{equation}
If this expression is evaluated for the choice of the Fisher Log-Series, it becomes
\begin{equation}
\displaystyle
A_1 = h(x) A_2 = (1-x) \frac{-x-\log(1-x)}{x^2} A_2 \ .
\label{suppeq:h-x}
\end{equation}

\section{Endemic Area Relationship}
\label{suppsec:EAR}

In this section, following the same procedure used to calculate the SAR, we obtain an expression for the
Endemic Area Relationship (EAR) in a large homogeneous system. The EAR is defined as the average number of species
whose population is completely contained in an area $A$.

Given a system of area $A_0$, the number of endemic species in an area $A$ will be equal to the number of
species, with at least one individual in $A_0$, which do not have an individual outside $A$ (i.e. in the area $A_0 \backslash A$).
We obtain a relation between the SAR and the endemic area relationship $E(A)$
\begin{equation}
\displaystyle
\begin{split}
& E(A|A_0) = S(A_0) - S(A_0 \backslash A|A_0) = \frac{S_{tot}(A_0)}{A_0} E(A) = \\
& \frac{S_{tot}(A_0)}{A_0}
\int_{A_0} d^2 \underline{z} \int_0^\infty d \lambda p(\lambda) e^{- \lambda} \big[
e^{ \lambda \int_{A(\underline{z})} d^2 \underline{r} \phi(\underline{r})  } -1 \big] \ ,
\end{split}
\label{suppeq:ear}
\end{equation}
which, in the continuous limit, becomes equation~\ref{eq:EAR-final} of the main text.

By using the same arguments used for the distribution of the number of species,
it is possible to demonstrate that
the probability to find $k$ endemic species in an area $A$ is a Poisson distribution of average $E(A)$, i.e.
\begin{equation}
\displaystyle
P_k^E(A) = \frac{ \big( E(A) \big)^k }{k!} \exp(-E(A)) \ .
\label{suppeq:EARprob}
\end{equation}


\section*{Acknowledgement}
J.G. thanks B.~Bassetti, M.~Cosentino~Lagomarsino and A.~Sanzeni for many useful discussions.
S.A. was supported by the EU FP7 SCALES project ("Securing the Conservation of biodiversity across Administrative Levels and spatial, temporal and Ecological Scales"; project No. 26852). A.M. thanks Cariparo foundation for financial support.
We thank S.J. Cornell and W.E. Kunin for insightful discussions.

\bibliographystyle{unsrt}
\bibliography{biblio}

\end{document}